\documentclass[prl,amssymb,twocolumn,tightenlines,showpacs]{revtex4-1}

\usepackage{graphicx}
\usepackage{amsmath}
\usepackage{verbatim}
\usepackage{subfigure}

\begin{document}

\title{Projection of two biphoton qutrits onto a maximally entangled state}

\author{A. Halevy}
\affiliation{Racah Institute of Physics, Hebrew University of
Jerusalem, Jerusalem 91904, Israel}
\author{E. Megidish}
\affiliation{Racah Institute of Physics, Hebrew University of
Jerusalem, Jerusalem 91904, Israel}
\author{T. Shacham}
\affiliation{Racah Institute of Physics, Hebrew University of
Jerusalem, Jerusalem 91904, Israel}
\author{L. Dovrat}
\affiliation{Racah Institute of Physics, Hebrew University of
Jerusalem, Jerusalem 91904, Israel}
\author{H. S. Eisenberg}
\affiliation{Racah Institute of Physics, Hebrew University of
Jerusalem, Jerusalem 91904, Israel}


\begin{abstract}
Bell state measurements, in which two quantum bits are projected
onto a maximally entangled state, are an essential component of
quantum information science. We propose and experimentally
demonstrate the projection of two quantum systems with three
states (qutrits) onto a generalized maximally entangled state.
Each qutrit is represented by the polarization of a pair of
indistinguishable photons - a biphoton. The projection is a joint
measurement on both biphotons using standard linear optics
elements. This demonstration enables the realization of quantum
information protocols with qutrits, such as teleportation and
entanglement swapping.
\end{abstract}
\maketitle

Quantum entanglement is a basic resource for quantum information
communication and processing protocols. It can be achieved either by
certain physical interactions between two quantum systems
\cite{Kwiat95} or by a projection of both systems with a specific
measurement \cite{Weinfurter94}. The projection of two independent
quantum states onto a maximally entangled state destroys them.
Nevertheless, such a projection is required for the implementation
of innovative quantum information protocols, such as quantum
teleportation \cite{Bennett93,Bouwmeester97} and entanglement
swapping \cite{Pan98}, which can be used to realize quantum
repeaters \cite{Briegel98}.

Quantum information is often encoded in the polarization of photons.
It is possible to realize a three-level quantum system, a qutrit, by
the polarization of a pair of photons
 \cite{Burlakov99}. If the two photons are indistinguishable in any
degree of freedom except for their polarization, they occupy the
three-dimensional symmetric triplet subspace of the Hilbert space
of two qubits. Such a general biphoton state can be written as:
\begin{equation}\label{GeneralQutrit}
|\psi_{bi}\rangle=(\alpha_0\frac{a_h^{\dagger2}}{\sqrt{2}}+\alpha_1a_h^{\dagger}a_v^{\dagger}+\alpha_2\frac{a_v^{\dagger2}}{\sqrt{2}})|\texttt{vac}\,\rangle\,,
\end{equation}
where $a_h^{\dagger}$ ($a_v^{\dagger}$) is the creation operator of
a horizontally (vertically) polarized single photon in the spatial
mode $a$, $\alpha_0$, $\alpha_1$, and $\alpha_2$ are the amplitudes
of the three biphoton states, and $|\texttt{vac}\,\rangle$ is the
vacuum state. Using linear optical elements limits the possible
rotations of biphoton qutrits \cite{Bregman08}. Nevertheless, adding
ancilla photons increases the amount of accessible rotations
\cite{Lanyon08}. Qutrits have also been realized by superimposing a
single photon between three orbital momentum
\cite{Mair01,Groblacher06}, spatial \cite{Bartuskova06}, and
temporal \cite{Thew04} modes. Arbitrary rotations can be applied to
such qutrit realizations \cite{Reck94}, and there is even a
suggestion for a conversion technique between biphotons and single
photons in three modes \cite{Lin09}. The incorporation of qutrits
can also enhance the performance of quantum communication protocols
\cite{Fitzi01,Cabello02,Brukner02,Langford04}. Recently, biphoton
qutrits have been shown to improve the security and efficiency of
quantum key distribution \cite{Bregman08}.

In the case of two qubits, there are four possible maximally
entangled states, known as the Bell states. Various photonic Bell
state projections have been realized by using linear optics elements
[beam-splitters, polarization beam-splitters (PBSs), and wave
plates] \cite{Mattle96,Giuseppe97}. However, linear optics enables
the simultaneous detection of only two of the four Bell states
\cite{Vaidman99,Lutkenhaus99}. A projection onto all four states has
been demonstrated by a nonlinear interaction with low efficiency
\cite{Kim01} and when hyper-entanglement is present, by using the
additional entangled degrees of freedom \cite{Kwiat98,Schuck06}.
Other demonstrations of full Bell state projection used auxiliary
photons \cite{Zhao05,Walther05} or path-entangled single photons
\cite{Boschi98}.

Single photons can represent quantum systems of $d$ states (qudits)
by occupying $d$ different modes
\cite{Mair01,Groblacher06,Bartuskova06,Thew04}. The projection of
two such photonic qudits of $d>2$ onto a maximally entangled state
is impossible without the use of auxiliary photons. This is because
only two particles are involved, while the Schmidt number of the
projected state is larger than 2 \cite{Calsamiglia02}. However, in
the case of two biphoton qutrits, four particles are involved and a
projection onto a maximally entangled qutrit state is possible.

In this Letter, we present an experimental scheme composed of linear
optical elements that can project two biphoton qutrits onto a
maximally entangled state. A generalized Bell basis of nine
maximally entangled states is presented. We explicitly show that our
scheme can discriminate one out of the nine states. A successful
projection results in a coincidence detection of all four photons
involved by single-photon detectors. Experimental results of such
successful projections are presented.

We designate the three biphoton vectors whose amplitudes in Eq.
\ref{GeneralQutrit} are $\alpha_0$, $\alpha_1$, and $\alpha_2$ with
the logical representation $|0\rangle$, $|1\rangle$ and $|2\rangle$,
respectively. One possible way to write the maximally entangled
basis of two qutrits is \cite{Bennett93}
\begin{equation}\label{GeneralPsi}
|\psi_{mn}\rangle=\frac{1}{\sqrt{3}}\sum_{j}\tau^{jn}|j\rangle_a\otimes|(j+m)\textsf{mod}\,3\rangle_b\,,
\end{equation}
where $\tau$ is defined as $e^\frac{2\pi i}{3}$ - the third root of
unity - $a$ and $b$ are spatial modes, and the values of $m$, $n$,
and $j$ can be 0, 1, and 2. From this general form, nine maximally
entangled biphoton states can be explicitly written, arranged in
three groups according to their $m$ values,
\begin{eqnarray}\label{3Families}
\nonumber|\psi_{0n}\rangle&=&\frac{1}{\sqrt{12}}(a_h^{\dag2}b_h^{\dagger2}+2\tau^{n} a_h^{\dagger}a_v^{\dagger}b_h^{\dagger}b_v^{\dagger}+\tau^{2n}a_v^{{\dagger}2}b_v^{{\dagger}2})|\texttt{vac}\,\rangle,\\
\nonumber|\psi_{1n}\rangle&=&\frac{1}{\sqrt{6}}(a_h^{\dagger2}b_h^{\dagger}b_v^{\dagger}+\tau^{n} a_h^{\dagger}a_v^{\dagger}b_v^{\dagger2}+\frac{1}{\sqrt{2}}\tau^{2n}a_v^{\dagger2}b_h^{\dagger2})|\texttt{vac}\,\rangle,\\
\nonumber|\psi_{2n}\rangle&=&\frac{1}{\sqrt{6}}(\frac{1}{\sqrt{2}}a_h^{\dagger2}b_v^{\dagger2}+\tau^{n} a_h^{\dagger}a_v^{\dagger}b_h^{\dagger2}+\tau^{2n}a_v^{\dagger2}b_h^{\dagger}b_v^{\dagger})|\texttt{vac}\,\rangle.\\
\end{eqnarray}
The three groups differ from each other by permutations of the $b$
mode vector. The three vectors of each group differ by the phase
difference between their terms. Here we focus on the detection of
states from the $|\psi_{0n}\rangle$ family.

If a certain configuration produces the qubit photonic Bell state
$|\phi^+\rangle$ \cite{Kwiat95}, its second-order process, i.e.,
when two photons of a single pulse split simultaneously, results in
the state \cite{Lamas-Linares01}
\begin{equation}\label{Phi2}
|\phi^{(2)}\rangle=\frac{1}{\sqrt{12}}(a_h^{\dagger2}b_h^{\dagger2}+2e^{i\delta}a_h^{\dagger}a_v^{\dagger}b_h^{\dagger}b_v^{\dagger}+e^{2i\delta}a_v^{\dagger2}b_v^{\dagger2})|\texttt{vac}\,\rangle\,.
\end{equation}
By tuning the angle $\delta$, by using a birefringent crystal in one
of the spatial modes, between $\delta=0^\circ$, $120^\circ$, and
$240^\circ$, it is possible to continuously change between the three
states $|\psi_{00}\rangle$, $|\psi_{01}\rangle$, and
$|\psi_{02}\rangle$, respectively.

\begin{figure}[tb]
\centering\includegraphics[angle=0,width=86mm]{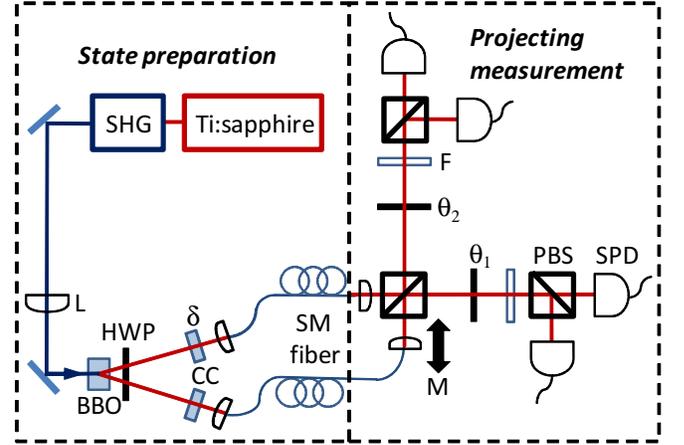}
\caption{\label{Fig1}(Color online) The experimental setup: a
mode-locked Ti:sapphire laser at 780\,nm is up-converted by second
harmonic generation (SHG). The 230\,mW beam is focused by a lens (L)
on a 2\,mm thick $\beta$-barium-borate (BBO) crystal, cut for the
generation of polarization entangled photon pairs via type-II
parametric down-conversion. The down-converted photons flip
polarization at a half-wave plate (HWP) at $45^\circ$ and temporal
walk-off is corrected by compensating crystals (CC). One of the
crystals is used to tune the birefringent angle $\delta$ [Eq.
\ref{Phi2}]. The photons are coupled into single mode fibers (SM)
where their polarization is adjusted by polarization controllers.
After the photons from both modes exit the fibers, they are
overlapped at a polarization beam-splitter. The relative propagation
delay between the two optical paths is adjusted by translating one
of the fiber ends by a linear motor (M). The two half-wave plates at
angles $\theta_1$ and $\theta_2$ are used to complete the projection
process. After passing the projection configuration, the photons are
spectrally filtered by using 3\,nm wide bandpass filters (F) and
coupled into multimode fibers that guide them to single-photon
detectors (SPD).}
\end{figure}

In order to explain our projection scheme, we first examine the
output of a PBS when each state of the three maximally entangled
groups serves as its input. The first three states
$|\psi_{0n}\rangle$ pass through the PBS without any change, up to a
birefringent phase from the reflection of the vertically polarized
photons. This phase can be compensated for either before or after
the PBS. On the other hand, the remaining six states
$|\psi_{1n}\rangle$ and $|\psi_{2n}\rangle$ are not preserved by the
PBS operation. None of them results in a biphoton at each of the
output ports. Thus, postselecting only cases when two photons exit
from each of the PBS ports already discriminates between the three
states of the first group and the six other states. In order to
discriminate between the different members of the
$|\psi_{0n}\rangle$ family and complete the projecting measurement,
we add to the output of each of the PBS ports a half-wave plate at
an angle $\theta$ with the vertical direction (see Fig. \ref{Fig1}).
Finally, the two polarizations of each mode are split by another PBS
into two single-photon detectors. Regardless of the wave-plate
angles, the first PBS certifies that a simultaneous fourfold
detection at all of the detectors can originate only from states of
the $|\psi_{0n}\rangle$ form. Propagating this general state through
the presented setup results in nine different quantum amplitudes,
one for each of the different options to have two photons at each of
the two arms. The amplitude of the
$a_h^{\dagger}a_v^{\dagger}b_h^{\dagger}b_v^{\dagger}$ term for
fourfold detection can be simplified to
\begin{equation}\label{4Fold}
A_{4f}=\frac{1}{\sqrt{12}}(\sin^24\theta+2\tau^{n}\cos^24\theta+\tau^{2n}\sin^24\theta)\,.
\end{equation}
The requirement that this amplitude should vanish when $n=1$ and 2
is fulfilled when
\begin{equation}\label{AngleCondition}
\tan^24\theta=2\longrightarrow\theta\simeq13.68^\circ\,.
\end{equation}

This result can be generalized by allowing the two wave-plate
angles to differ from each other. It can be shown that the
condition for two independent wave-plate angles $\theta_1$ and
$\theta_2$ is
\begin{equation}\label{2AnglesCondition}
\tan4\theta_1 \tan4\theta_2=2\,.
\end{equation}
Nevertheless, the maximum theoretical fourfold coincidence rate is
obtained in the degenerate case. Thus, by setting the angles of the
two wave plates to $13.68^\circ$, fourfold coincidence can originate
only from $|\psi_{00}\rangle$ with a detection probability of
$\frac{1}{3}$. The probability is smaller than 1 as the
$|\psi_{00}\rangle$ state can also result in quantum states other
than $a_h^{\dagger}a_v^{\dagger}b_h^{\dagger}b_v^{\dagger}$. The
source of those other states cannot be distinctively related to a
single state, and thus their detection is useless for entanglement
projection. The setup can be altered to detect the
$|\psi_{01}\rangle$ or the $|\psi_{02}\rangle$ states as well, by
adding a birefringent phase in one arm before the PBS of
$-120^\circ$ or $-240^\circ$, respectively. Interestingly, the
fourfold detection amplitude vanishes when one wave plate is at
$0^\circ$ and the other at $22.5^\circ$ for all $|\psi_{0n}\rangle$
states. This parameter setting was recognized in a previous work in
the context of nonlocal bunching \cite{Eisenberg05}.

The discussion until now assumed complete spatiotemporal overlap
between the photons in spatial modes $a$ and $b$. When the optical
paths leading to the PBS differ, the photons in each mode are
temporally labeled and therefore are distinguishable. The fourfold
coincidence term
$a_h^{\dagger}a_v^{\dagger}b_h^{\dagger}b_v^{\dagger}$ is split into
six noninterfering terms. The expected fourfold coincidence
detection probability for each of the three $|\psi_{0n}\rangle$
states becomes $\frac{2}{9}$. Thus, we expect all three states to
have the same background rate when temporal distinguishability is
introduced, as this is a projection onto a uniform mixture of these
states. When distinguishability is removed, the rates from states
$|\psi_{01}\rangle$ and $|\psi_{02}\rangle$ decrease to zero while
the $|\psi_{00}\rangle$ rate would have a 50$\%$ increase over the
background.

\begin{figure}[tb]
\centering\includegraphics[angle=0,width=86mm]{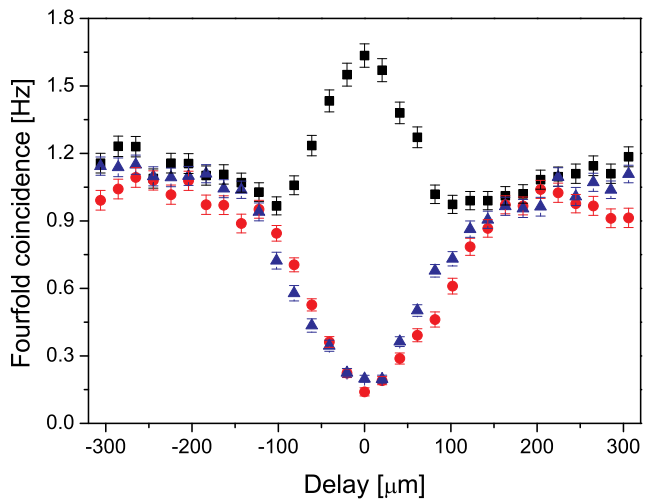}
\caption{\label{Fig2}(Color online) Experimental results of the
fourfold coincidence rates as a function of the relative delay for
the three $|\psi_{0n}\rangle$ states. Each data point was averaged
over 600 seconds. The desired $|\psi_{00}\rangle$ state (black
squares) shows a relative increase in the coincidence rate by
about 50\% at zero delay, when compared to large delays. The
$|\psi_{01}\rangle$ (red circles) and $|\psi_{02}\rangle$ (blue
triangles) states are well rejected, as shown by their
considerable signal decrease at zero delay.}
\end{figure}

To demonstrate the suggested projection, we used the experimental
setup shown in Fig. \ref{Fig1}. When the half wave-plate angles
$\theta_1$ and $\theta_2$ are both set at $22.5^\circ$, this setup
becomes identical to a two qubits $|\phi^{\pm}\rangle$ projecting
setup. We used this configuration to calibrate the birefringent
angle $\delta$ as well as to find the zero delay position. A total
rate of about 25000 twofold coincidences per second was observed
with a dip visibility of $(95\pm1)\%$ (not presented here). The high
twofold visibility figure indicates the quality of the PBS
projection \cite{Poh09}.

The states of the $|\psi_{1n}\rangle$ and $|\psi_{2n}\rangle$
families can not be generated by our setup. Nevertheless, their
rejection is a result of the basic nature of PBSs. The PBS we have
used has 5\% reflection for a horizontally polarized beam, whereas
an ideal PBS should have none. Using this figure, we calculated the
probability to detect a state from these families to be around 1\%
in the worst possible case, when the input is a mixture of these
families. Thus, we concentrate on demonstrating the discrimination
between the three states of the $|\psi_{0n}\rangle$ family. Using
the angle calibration, we tuned $\delta$ to be $0^\circ$,
$120^\circ$, and $240^\circ$ in order to produce the
$|\psi_{00}\rangle$, $|\psi_{01}\rangle$, and $|\psi_{02}\rangle$
states, respectively. The wave-plate angles were tuned to
$13.68^\circ$ according to Eq. \ref{AngleCondition}, and the delay
between the paths was scanned. We present the results of these scans
in Fig. \ref{Fig2}. At the region of no temporal overlap (large
delay), the three states share the same fourfold event background
rate of $1.1\pm0.1$ Hz. When distinguishability is removed at the
zero delay position, $|\psi_{00}\rangle$ shows a clear rise in the
fourfold rate to a maximal value of $1.63\pm0.05$ Hz, while the
orthogonal states $|\psi_{01}\rangle$ and $|\psi_{02}\rangle$
decrease to minimal values of $0.14\pm0.02$ and $0.19\pm0.02$ Hz,
respectively. The background rates at zero delay for
$|\psi_{01}\rangle$ and $|\psi_{02}\rangle$ are attributed to both
imperfect initial state generation and the higher order term of six
photons. The ratio between the projected and the rejected states is
about 10, indicating a good projection fidelity. The ratio between
the $|\psi_{00}\rangle$ signal values at zero delay and at large
delay is $\sim1.5$, in accordance with the theoretical prediction up
to experimental error. As the theoretical maximum efficiency of this
projection is $\frac{1}{3}$, these rates indicate a combined
production and detection rate of about 5 fourfold events per second.

\begin{figure}[tb]
\centering\includegraphics[angle=0,width=86mm]{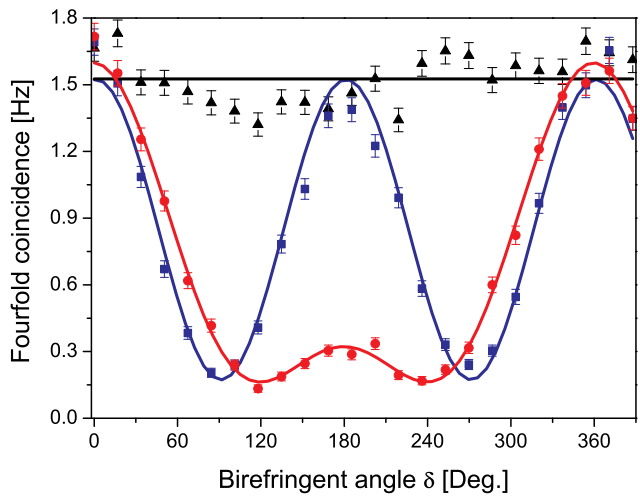}
\caption{\label{Fig3}(Color online) Experimental results of the
fourfold coincidence rates as a function of the birefringent angle
$\delta$ for three wave-plate angle settings: $\theta=0^{\circ}$
(black triangles), $\theta=22.5^{\circ}$ (blue squares) and
$\theta=13.68^{\circ}$ (red circles). Each data point was averaged
over 480 seconds. The solid lines represent the fit to theoretical
predictions.}
\end{figure}

As the birefringent angle $\delta$ is continuous, it is possible to
generate not just the three $|\psi_{0n}\rangle$ states but a
continuous range, as can be seen by Eq. \ref{Phi2}. We set the delay
position to zero, where projection occurs, and scanned this angle.
The expected rate is $|A_{4f}|^2$ [Eq. \ref{4Fold}], where we
replaced the discrete variable $n$ with a continuous phase
$\delta=\frac{2\pi}{3}n$. The results of this measurement are shown
in Fig. \ref{Fig3}. Measurements for three wave-plate settings were
performed. At $\theta=0^\circ$ only the central term of the rate
[Eq. \ref{4Fold}] is nonzero, and thus when calculating
$|A_{4f}|^2$, there is no interference between the three terms and
the rate is independent of $\delta$. When the wave plates' angle is
$\theta=22.5^\circ$, the central term is always zero and the first
and last terms have identical magnitudes. These two terms interfere
as $\cos^2\delta$. The third wave-plate setting is at
$\theta=13.68^\circ$, where all three terms interfere with the same
magnitude. As expected, this curve has a maximum value at
$\delta=0^\circ$ (corresponding to the $|\psi_{00}\rangle$ state)
and two minima at $\delta=120^\circ$ and $240^\circ$, when the
states $|\psi_{01}\rangle$ and $|\psi_{02}\rangle$ are generated.
The experimental data fit the predicted theoretical curves well. The
visibilities of the $\theta=22.5^\circ$ and $\theta=13.68^\circ$
curves are $0.77\pm0.1$ and $0.82\pm0.04$, respectively. Deviations
(most apparent in the $\theta=0^\circ$ data) are mainly due to laser
power and coupling fluctuations during the long overall acquisition
time of about 11 hours.

A quantum communication network with qutrits will provide many
advantages that were suggested in previous works
\cite{Bregman08,Fitzi01,Cabello02,Brukner02,Langford04}. For such
a network to be practical, qutrit quantum repeaters should be
available. These future devices are based on entanglement swapping
which in turn relies on projection onto maximally entangled
states. As mentioned above, it is impossible to realize such a
projection with linear operations on qutrits that are represented
by single photons \cite{Calsamiglia02}. This fact rules out as
practical candidates most of the qutrit realizations to date.
Polarized biphotons are a clear exception to this, as is shown by
this work.

In a recent work, two photons from two different entangled pairs
have been projected onto the symmetric triplet subspace of biphotons
\cite{Kaltenbaek10}. This projection created a qubit-qutrit-qubit
chain, which is a valence-bond solid ground state \cite{Darmawan10}.
However, since the projection involves only two single photons, it
is equivalent to the standard qubit Bell measurement
\cite{Weinfurter94}.

In conclusion, we have demonstrated a linear optics scheme that
projects two biphoton qutrits onto a maximally entangled state.
Three such orthogonal states were prepared and the scheme's
ability to discriminate between them has been shown. When a
continuous phase between these states was scanned, the output
fourfold signal has been shown to fit the theoretical prediction
very well. The demonstration of qutrit teleportation using this
projection is currently under progress.

The authors thank the Israeli Science Foundation for supporting this
work under Grants No. 366/06 and No. 546/10.

\end{document}